\documentclass[sigconf,10pt]{acmart}
\makeatletter
\def\@ACM@checkaffil{
    \if@ACM@instpresent\else
    \ClassWarningNoLine{\@classname}{No institution present for an affiliation}%
    \fi
    \if@ACM@citypresent\else
    \ClassWarningNoLine{\@classname}{No city present for an affiliation}%
    \fi
    \if@ACM@countrypresent\else
        \ClassWarningNoLine{\@classname}{No country present for an affiliation}%
    \fi
}
\makeatother

\usepackage{makecell}
\usepackage{indentfirst}
\usepackage{booktabs} 
\usepackage{enumerate} 
\usepackage{stfloats}
\usepackage{natbib}
\usepackage{verbatim}
\usepackage{url}
\usepackage{times}
\usepackage{soul}
\usepackage{listings}
\usepackage{algorithm}
\usepackage{algorithmic}
\usepackage[algo2e]{algorithm2e} 
\usepackage{xcolor}
\usepackage{color}
\usepackage{mathrsfs}  
\usepackage{multirow}  
\usepackage{subfigure}
\usepackage{graphicx}
\usepackage[normalem]{ulem}

\renewcommand\footnotetextcopyrightpermission[1]{} 
\newcommand\sysname{Miriam}
\title{\huge Miriam: Exploiting Elastic Kernels for Real-time Multi-DNN Inference on Edge GPU}
\author{Zhihe Zhao$^{\dagger}$, Neiwen Ling$^{\dagger}$, Nan Guan$^{\S}$, and Guoliang Xing$^{\dagger}$}
\affiliation{\institution{$^{\dagger}$The Chinese University of Hong Kong} }
\affiliation{\institution{$^{\S}$City University of Hong Kong}}

\usepackage{xcolor}

\usepackage{stfloats}

\definecolor{codegreen}{rgb}{0,0.6,0}
\definecolor{codegray}{rgb}{0.5,0.5,0.5}
\definecolor{codepurple}{rgb}{0.58,0,0.82}
\definecolor{backcolour}{rgb}{0.95,0.95,0.92}

\definecolor{C1}{RGB}{255,255,255}
\definecolor{C2}{RGB}{0,102,204}

\lstdefinestyle{mystyle}{
    backgroundcolor=\color{C1},   
    commentstyle=\color{C2},
    keywordstyle=\color{magenta},
    numberstyle=\tiny\color{codegray},
    stringstyle=\color{codepurple},
    basicstyle=\ttfamily\footnotesize,
    breakatwhitespace=false,      
    morekeywords={
        cudaMalloc, cudaFree,
        __global__, __shared__, __device__, __host__,
        __syncthreads,
    },
    breaklines=true,                 
    captionpos=b,                    
    keepspaces=true,                 
    numbers=left,                    
    numbersep=5pt,                  
    showspaces=false,                
    showstringspaces=false,
    showtabs=false,                  
    tabsize=2
}

\begin{document}
\fancyhead{}


\begin{abstract}
\noindent Many applications such as autonomous driving and augmented reality, require the concurrent running of multiple deep neural networks (DNN) that poses different levels of real-time performance requirements. However, coordinating multiple DNN tasks with varying levels of criticality on edge GPUs remains an area of limited study. Unlike server-level GPUs, edge GPUs are resource-limited and lack hardware-level resource management mechanisms for avoiding resource contention. Therefore, we propose Miriam, a contention-aware task coordination framework for multi-DNN inference on edge GPU. Miriam consolidates two main components, an elastic-kernel generator, and a runtime dynamic kernel coordinator, to support mixed critical DNN inference. To evaluate Miriam, we build a new DNN inference benchmark based on CUDA with diverse representative DNN workloads. Experiments on two edge GPU platforms show that Miriam can increase system throughput by 92\% while only incurring less than 10\% latency overhead for critical tasks, compared to state of art baselines.

\end{abstract}
 
\maketitle
\section{Introduction} \label{intro}
Deep learning (DL) has become a catalyst for a wide range of applications running on the edge, such as augmented reality and autonomous driving. These applications typically require the concurrent execution of multiple DNN tasks that have varying levels of criticality. For example, in mobile augmented reality, DNN inference tasks are often used for gesture recognition and user behaviour analysis, which are key components in providing a seamless user experience. This presents a major challenge as mobile/edge devices are constrained by limited computational resources for running multi-DNN inference tasks in real-time.


To support multiple DNN-based applications that have different real-time requirements \cite{reef}, a common practice is to share an edge Graphics Processing Unit (GPU). However, this practice poses significant challenges. On the one hand, when executing multiple DNNs simultaneously, their contention over the limited onboard resources on the same edge GPU can result in a performance bottleneck \cite{interference}. 
On the other hand, dedicating the entire GPU to latency-critical tasks to guarantee their real-time requirements results in low GPU utilization \cite{automatediccad}. Meanwhile, most of the approaches that attempt to support concurrent DNN inference tasks on GPU \cite{warpslicer,smcentric,dynamicSchedulueGPU} require runtime support from vendors like NVIDIA Multi-Process Service (MPS) and Multi-Instance GPU (MIG) \cite{MIG,MPS}, which are unavailable on edge GPUs due to the architectural differences.


Furthermore, multi-DNN inferences present two potentially conflicting objectives. Firstly, it is imperative that critical DNN tasks are given priority over other tasks in order to minimize end-to-end latency. This necessitates that the critical tasks are treated as first-class citizens on the GPU, with no interference from other tasks. Secondly, in order to achieve high overall throughput, all co-running DNN tasks should be concurrently executed in a best effort manner. These two conflicting objectives pose a major challenge for efficiently coordinating the inferences of multiple DNN tasks on edge GPU.

In this paper, we propose a new system named Miriam which aims to support real-time multi-DNN inference on edge GPUs by addressing the latency and throughput problems of co-running multiple DNN inference tasks. \textit{The key idea of Miriam is based on the elastic kernel \footnote{Kernel here refers to a small program that is executed on a GPU to perform the specific DNN kernel computations.}, which can achieve more fine-grained resource mappings on GPU.} Specifically, traditional kernels are elasticized by breaking them down into smaller, more flexible units that can be dynamically scheduled and remapped to different GPU resources based on their priority and criticality. This elasticization approach enables the padding of other GPU kernels, which maximizes GPU utilization without causing significant resource contention. As a result, critical tasks can be prioritized without compromising overall system throughput, thus improving the real-time performance of the system.

Our design is based on the key observation that the latency degradation of co-running DNN kernels is mainly caused by two dominant factors, namely \textit{intra-multi-processor (SM) resource contention} and \textit{inter-multi-processor resource contention}. 
We leverage elastic kernels to address those two kinds of resource contention. Specifically, Miriam integrates two main components. The first component, the \textit{elastic-kernel generator}, consists of an elastic grid/block generator that generates resource-controllable GPU kernels to resolve co-running DNN tasks resource contention, and a source-to-source kernel transformer that converts original GPU kernels into elastic kernels while preserving computation consistency. We also design a \textit{dynamic runtime coordinator} to schedule the elastic kernels to proactively control the execution of the co-running kernel at runtime.
To evaluate the effectiveness of Miriam, we implement it as a hybrid framework based on CUDA, C++, and Python. We use a set of multi-DNN inference benchmarks for edge GPUs that include tasks with different priorities to evaluate the system's effectiveness. Our results demonstrate that, compared to existing methods, Miriam can serve significantly more requests with up to 92\% throughput improvement while maintaining the inference speed for critical tasks with only a 10\% increase in latency. These results highlight Miriam's superior performance in achieving efficient coordination of real-time multi-DNN inference tasks on edge GPUs.

\vspace{-1em}
\section{Related Work} \label{related}
To enable on-device multi-DNN inference on edge devices, prior methods such as joint DNN model compression sacrifices a modest level of accuracy for each model to reduce the computational costs of mixed DNN workloads \cite{rt-mdl,nestdnn, deepeye}. In contrast, Miriam does not compromise on accuracy and can be seen as an orthogonal approach to the above systems. Other methods address this problem through new compiling techniques. For example, Veltair \cite{veltair} proposes to generate multiple versions of compiled DNN models with different intensities of resource contention for scheduling at runtime to accelerate multi-DNN inference. However, these methods also lead to issues such as high overhead in storage and offline profiling, making them hard to scale to more use cases. 

Systems like DeepEye \cite{deepeye}, Abacus \cite{enablesc}, and Dart \cite{dart} have utilized the interleaving of operators with different "contention channels" (memory-bound or compute-bound). Although these methods have proven to be effective, they require time-consuming offline profiling and are cumbersome to generalize for new DNN tasks. REEF \cite{reef} addresses the same problem of mixed-critical multi-DNN inference coordination and achieves kernel-level preemption for critical tasks. However, the approach requires modification of the GPU driver library, which is not practical in many popular closed-source devices. Heimdall \cite{yi2020heimdall} and Band \cite{band} also target solving resource contention of multi-DNN inference, while they have different application settings from ours.

Warped-Slicer \cite{warpslicer} employs performance versus computing unit occupancy curves for selecting an optimized simultaneous kernel pattern, but the method fails to address resource contention between kernels. Works such as HSM \cite{hsm} and \cite{wang2019efficient} model the latency degradation of concurrent GPU kernel executions based on hardware information, but the predictors built in these works are difficult to adapt to real-world multi-DNN inference scenarios that are characterized by nondeterministic kernel overlapping \cite{enablesc}. Other works such as Smcentric \cite{smcentric} and Effisha \cite{effisha} tackle the GPU multitasking problem from resource management perspectives in a space-multiplexing manner \cite{wu2020model,jain2019fractional}, which is orthogonal to Miriam's approach.


\begin{figure}[t!]
\centering
\includegraphics[width=0.25\textwidth]{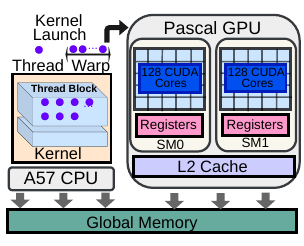}
\caption{An overview of CUDA programming paradigm and the computation hardware in NVIDIA TX2. }
\label{1}
\end{figure}

\section{Background}\label{background}
In this paper, we present the design and implementation of Miriam based on the CUDA programming model for NVIDIA GPU \cite{cuda}. We first introduce some terminologies in CUDA. Fig. \ref{1} (left) shows the layout of an NVIDIA Jetson TX2 GPU, which consists of two SMs, each capable of running a number of GPU threads with a maximum size, and both SMs share the global memory.

\textbf{CUDA Programming Model.} A CUDA GPU has a number of \emph{Streaming Multiprocessor (SM)}. Each SM contains multiple cores, which are the processing units that execute the instructions of the threads. All cores within the same SM share the same set of registers and can communicate with each other through shared memory.
Code executed by the GPU is known as a GPU \emph{kernel} \cite{demystifying}.
\emph{Threads} are the smallest unit of work that can be executed in parallel on a GPU, and they are organized into \emph{blocks}. Each block is a group of threads that can execute concurrently on a single SM. 
A \emph{grid} is a collection of blocks that are organized in a three-dimensional array. 
The grid defines the overall structure of the data being processed and how it is partitioned into blocks.
GPU \emph{streams} are a way of organizing and executing asynchronous tasks on the GPU. Each stream is a sequence of kernels (e.g. Conv, MemCopy) that can be executed independently of other streams. Kernels in the same stream are executed in a  FIFO manner \cite{cuda}. 

\textbf{Kernel Execution on GPU.} When launching a kernel in CUDA, we specify the dimensions of the grid and blocks. Each block is dispatched to and executed on one SM. However, whether a block can be dispatched to an SM that already has a block executing on it depends on whether there are enough remaining resources, such as thread slots and shared memory, to accommodate the new block. If there is no available SM to accommodate a block, it has to wait in a queue in a first-in, first-out (FIFO) order.
When a kernel executes on an SM, it competes for on-SM resources, such as thread slots and shared memory, with other kernels already dispatched to and executing on the same SM. This competition greatly affects the execution time of a kernel on the SM. Thus, the varying time a block waits in the queue, in addition to the varying time it takes to execute its workload on the SM, contributes to the overall varying latency experienced by the kernel.





\section{Motivation and Challenges}\label{challenges}
Miriam aims to support co-running DNN inference tasks on edge GPU for real-time applications. Tasks that have strict real-time requirements are referred to as critical tasks. For example, obstacle detection in autonomous driving must be finished by a certain deadline, allowing sufficient time for the vehicle to maneuver around obstructions. Tasks that do not have strict real-time deadlines are referred to as normal tasks. For example, monitoring human drivers' emotions and fatigue can be executed in a best-effort manner to improve the driving experience.

\sysname{} aims to meet the real-time requirement for latency-critical tasks while maximizing the overall throughput of co-running normal tasks in a dynamic manner. One common solution is to sequentially execute critical tasks and normal tasks, which can yield the lowest latency for critical task execution, but at the cost of significantly reduced overall throughput. An alternative solution is to directly execute multiple DNN tasks on the same edge GPU without proper contention management. However, this can cause increased latency for critical tasks. 

\begin{figure}[t!]
\centering
\begin{minipage}[t]{0.23\textwidth}
\centering
\includegraphics[width=4.2cm]{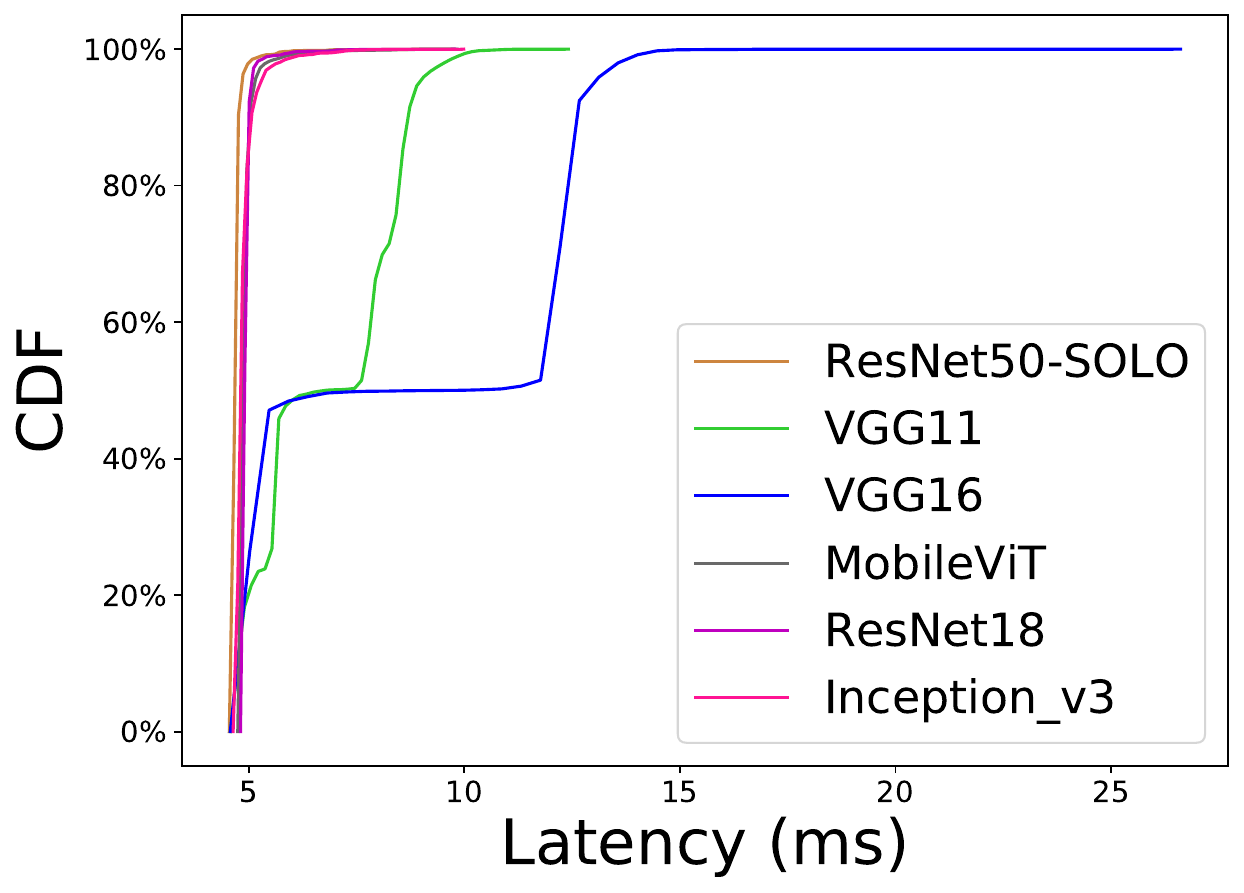}
\end{minipage}
\begin{minipage}[t]{0.23\textwidth}
\centering
\includegraphics[width=4.15cm]{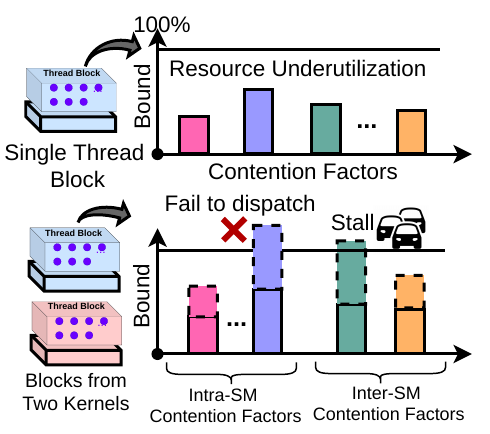}

\end{minipage}
\caption{(left) The latency distribution of ResNet50 when co-running with other DNN models. (right) Illustration of intra-SM and inter-SM contention.}
\label{moti3}
\end{figure}

Here we investigate performance degradation caused by the simultaneous execution of multiple DNN tasks. When running alone on an edge GPU, GPU kernel execution time for DNN inferences tends to remain consistent. However, the simultaneous execution of multiple DNN tasks on an edge GPU can significantly impact performance. To study this effect, we conducted an experiment using CUDA multi-stream on an NVIDIA RTX 2060 GPU where we launched a DNN task (i.e., ResNet50) with different co-runners in a closed-loop manner. In Fig.~\ref{moti3} (left), we present the cumulative distribution function (CDF) of the ResNet50 latency with various co-running tasks. The results show that the latency of ResNet50 ranges from 4.4 ms to roughly 16.2 ms when co-running with VGG16, while the solo-running latency is 4.2 ms, yielding a significant variation. Meanwhile, the latency distribution pattern for different co-running model settings also varies a lot.

The primary factor that results in these large variations in latency is the complex resource contention among the co-running tasks, which can be classified into \textit{intra-SM contention} and \textit{inter-SM contention}, as is shown in Fig.~\ref{moti3} (right). The latency experienced by a GPU kernel depends not only on the time it takes for the workload to execute on the SM (affected by intra-SM contention) but also on the time it takes for the workload to wait to be dispatched to the SM (affected by inter-SM contention). Intra-SM contention and inter-SM contention are two types of resource contention among co-running tasks on a GPU. Intra-SM contention refers to the contention within an SM, which can occur when multiple thread blocks from different kernels are dispatched to the same SM and compete for shared resources, such as registers, shared memory, and execution units. Inter-SM contention refers to the contention among SMs, which can occur when multiple thread blocks from different kernels are dispatched to different SMs and compete for shared resources, such as global memory and memory controllers. These two types of contention can cause significant performance degradation and latency variation for co-running tasks on a GPU. 

Thus, given two incoming DNN task queues for \textit{normal task $\tau^{normal{}}$} and \textit{critical task $\tau^{critical{}}$}, to maximize the overall task throughput while guaranteeing the real-time performance of critical tasks, it is crucial to carefully manage the contention that arises from multiple overlapping kernels during co-execution. Our design objective is: to mitigate the latency degradation of the critical kernel during concurrent execution with the normal kernel by resolving inter- and intra-SM contention while allocating idle SM resources to the normal kernel as much as possible.

\section{Miriam Overview} \label{system}

\begin{figure}[t!]
\centering
\includegraphics[width=0.45\textwidth]{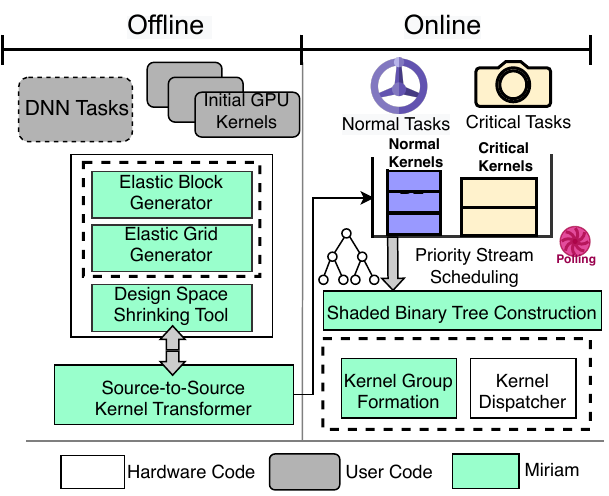}
\caption{Architecture of \sysname{}. Green boxes are our contributions. Modules in boxes with dashed border on the right are on the crucial route for handling DNN inference requests. The source-to-source kernel transformer module enables the elastic kernel design without incurring computation inconsistency.}
\label{ar}
\end{figure}
We now introduce Miriam, a holistic kernel-level system for real-time multi-DNN inference on edge GPU. Miriam is a compiler-runtime synergistic framework that achieves fine-grained kernel-level GPU resources mapping. In this section, we first introduce the key idea of Miriam and then describe its system architecture.



\vspace{-1em}
\subsection{Key Idea}
In Section \ref{challenges}, we show that it is imperative to give careful consideration to the resource contention that arises between multiple parallel kernels. Failure to do so can result in GPU under-utilization and degradation of inference latency.

Motivated by these findings, Miriam proposes a new DNN kernel inference abstraction, \emph{elastic kernel}, which is a GPU kernel that has adjustable grid size and block size. Different gird/block sizes of the elastic kernel correspond to different patterns of SM-level GPU resource usage. By transforming normal kernels into elastic kernels, Miriam can control their resource contention to the critical task, and thus maximize the overall system throughput while not compromising the real-time performance of the critical kernel. 


To this end, Miriam generates an elastic kernel for each normal task offline and enables kernel coordination at runtime. Specifically, Miriam employs a novel elastic kernel generator to construct an elastic kernel with adjustable GPU resource usage patterns. During the runtime phase, the coordinator will select the best implementation patterns of the elastic kernels and dynamically pad them with the critical kernels to fully utilize the GPU resource.
\vspace{-1em}
\subsection{System Architecture}
Fig.~\ref{ar} shows a bird-eye view of \sysname{}. Miriam incorporates two parts: \textit{Offline Elastic Kernel Generation} and \textit{Online Kernel Coordination}, working at levels of compilation, i.e., source-to-source code transformation, and kernel coordination, respectively. They collaborate to exploit elastic kernels for supporting multiple DNN inference on edge GPUs.
\sysname{} generates elastic kernels by transforming the compiler-generated or handcrafted CUDA kernels to the elastic form. We generate elastic kernels from both grids' and blocks' perspectives of GPU kernels, which are called elastic grid and elastic block, respectively. These configuration knobs can achieve fine-grained control over  inter- and intra-SM resources. 
\begin{figure}[t!]
\centering
\includegraphics[width=0.47\textwidth]{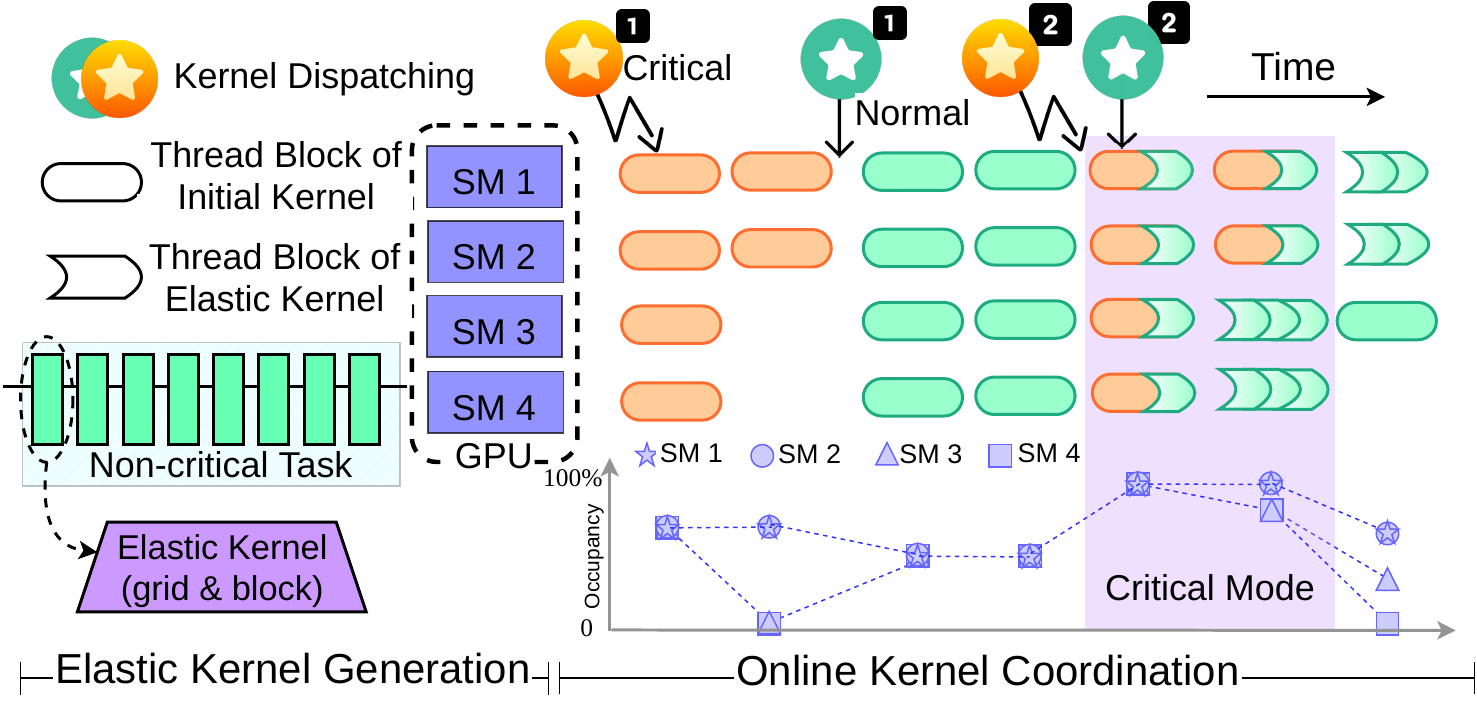}
\caption{An example of timeline in Miriam. The yellow star represents the critical kernel, while the green one represents the normal kernel.}
\label{timeline}
\end{figure}

There are two challenges here for generating elastic kernels. First, the design space of the elastic kernel implementation patterns is too large (e.g., 2874 on average for a single kernel in AlexNet \cite{tango}).
Hence, we shrink the design space to decrease the number of potential elastic kernel candidates by taking the hardware limitation into consideration. Second, When a kernel is launched in CUDA, the execution configuration specifies the number of threads to be launched and how they are organized into blocks and grids. Modifying the grid and block size in a DNN kernel directly can cause computation errors because this affects how threads are organized and executed on the GPU. In case of this, \sysname{} includes a novel \textit{source-to-source kernel transformer}, which transforms GPU programs of a given DNN kernel into an elastic kernel execution paradigm while ensuring the consistency of computation results. 



\sysname{} adopts a novel dynamic kernel coordination mechanism that controls the execution of elastic and critical kernels at run-time. Specifically, \sysname{} will profile the SM occupancy of each elastic kernel and the critical kernels. Then, \sysname{} determines the grid size and block size of the next {elastic kernel} from the normal task queue at runtime. In this way, tasks with elastic kernels can maximize resource utilization without interference to other co-running critical kernels.
{A key challenge here is that an elastic kernel may be executed solely or in parallel with different critical kernels. Hence, we cannot determine the scheduling of the elastic kernel at the time of kernel launch. To address this issue, we design a dynamic kernel sharding mechanism, in which we divide an elastic kernel into several shards and determine the scheduling for each sharding according to run-time resource usage.}



Miriam can support a wide range of applications that need to run multiple DNNs on the edge GPU.
For instance, an obstacle detection task and a navigation task need to run in parallel to achieve autonomous driving. 
The obstacle detection task is critical because it is related to driving safety, while the navigation task can be executed in a best-effort manner as a normal task.
For such a DL task set, as shown in Fig. \ref{timeline}, \sysname{} first divides them into critical kernels and normal kernels according to their task characteristic, i.e., criticality of the tasks. Normal kernels are compiled offline and transformed into elastic kernels by \sysname{}. At run-time, the elastic sharding policy of normal kernels is determined by the \sysname{} to maximize resource utilization while not interfering with the execution of the critical kernel.

\vspace{-1em}
\section{Generation of Elastic Kernels} \label{mgen}

\begin{figure}[t!]
\centering
\includegraphics[width=0.48\textwidth]{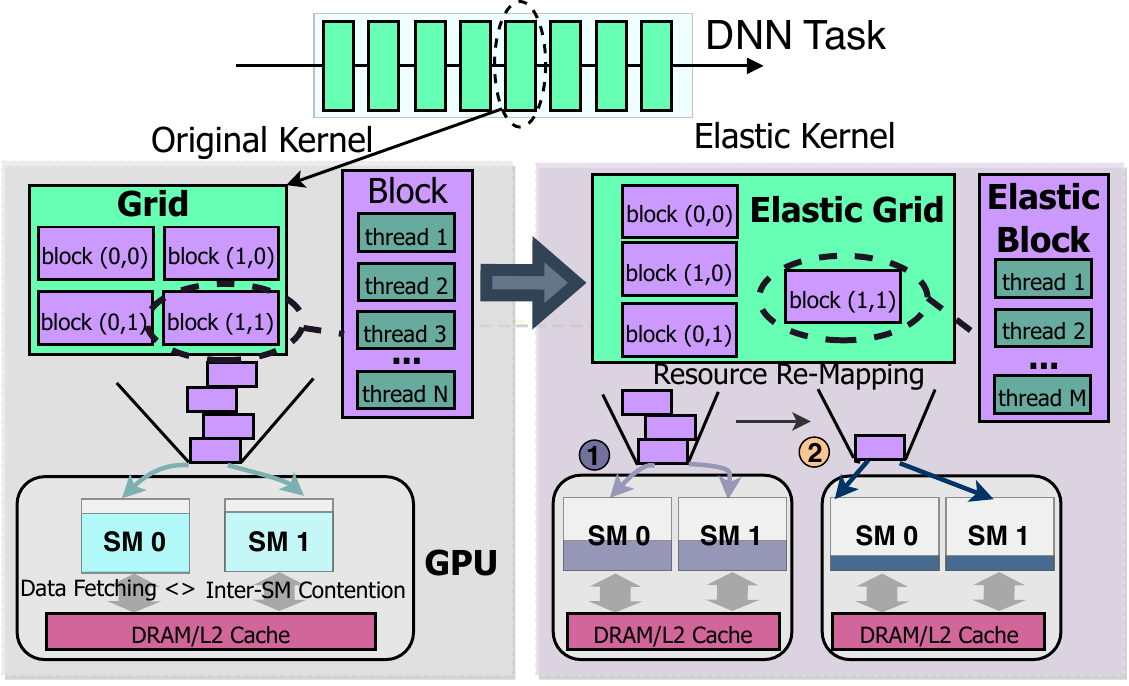}
\caption{Elastic kernel generation. The flexible re-mapping of SM-level GPU resources assisted by the elastic kernel generator enables the adaptation of the runtime dynamics in available resources.}
\label{elastic}
\end{figure}

To support finer control over inter- and intra-SM resources of a kernel running on the edge GPU, we propose an elastic kernel generator. The design principle of Miriam is based on the insight that both the block and grid's resource allocations can be distilled from the native GPU programming model. Fig. \ref{elastic} illustrates the design of the proposed elastic kernel generator: elastic block and elastic grid. By separating resource allocation for thread blocks from the logic-level grid and thread block identity, this approach generates resource-controllable GPU kernels for further resolving co-running DNN tasks resource contention problems.

To improve the efficiency of the elastic kernel generation process, \sysname{} proposes to shrink the design space of elastic kernels according to hardware limitations, as well as observations on co-running DNN kernels from critical and normal task queues. Moreover, to maintain the accuracy of elastic kernel calculation after elastic kernel transformation, we design a source-to-source kernel transformer. Our transformer can convert original GPU kernels into elastic kernels while preserving computational equivalence.



\subsection{Controllable Intra-SM Resource by Elastic Block} 



DNN kernels can be broadly categorized into memory operations (memory allocations, memory transfers, etc.) and kernel execution. To enable the execution of a single kernel on multiple GPU SMs, GPU programming divides a large kernel into multiple sub-kernels, each of which is executed by a GPU block. 
The block size is determined by the computation workload of each sub-computation. Blocks with smaller sizes consume less thread usage for each instruction cycle. 

Multi-DNN inference on edge GPU can cause severe intra-SM contention when multiple thread blocks from different kernels compete for the resource within the same SM. Some blocks would fail to execute or delay, which leads to a decrease in the overall throughput and an increase in the corresponding latency of the DNN inference. For this issue, one possible solution is to perform code-level optimization of the GPU kernel. This approach includes optimizing the memory access patterns and reducing unnecessary computations to decrease the intra-SM resource usage, and thus alleviates intra-SM contention. However, optimizing GPU codes for a specific DNN model is challenging and time-consuming. Different optimization techniques such as loop-tiling, loop-unrolling and parallelization naturally have different trade-offs in terms of execution performance, memory usage, and code complexity. Achieving the appropriate balance among those factors requires careful experimentation and tuning. 
Adapting codes for different concurrent kernels from diverse tasks demands a significant amount of effort and may not generalize well, thereby restricting the effectiveness and applicability of the optimization techniques.

To carefully manage the resource usage of each block, \sysname{} adjusts the number of threads within the targeted block to generate elastic blocks for each thread block. We adopt the persistent thread technique \cite{pt} that is capable of adjusting a kernel’s resident block size on an SM. In contrast to traditional kernels where threads terminate after completing the kernel execution, persistent threads remain active throughout the execution of a kernel function. We limit the range of each elastic block size to fall between 1 and the maximum resident block size. We also transform the default 1:1 logical-to-physical threads mapping scheme to an N:1 mapping scheme while preserving the initial program semantics.

Compared to static block fusion \cite{blockfu}, which fuses multiple thread blocks from different GPU kernels into a single one to reduce unnecessary loads and stores, our persistent thread design does not require pre-compilation of all possible combinations of kernels. This feature enables flexible SM-level resource mapping at runtime.

Our elastic kernel is designed to stay within the shared memory limit, and we achieve this by modifying the way we control the intra-SM resources, including shared memory, compared to the original kernel. This modification results in a memory occupancy that is either equal to or less than that of the original kernel.

While the persistent thread mechanism provides fine-grained control over intra-SM parallelism, it comes with nontrivial overhead. The optimal number of launched persistent threads does not always equal to the maximum number of concurrently executing threads from all thread blocks that can be afforded by a single SM. Hence, we will narrow the design space of elastic block which will be introduced in Section~\ref{subsec:deisgnspace}.


\subsection{Elastic Grid for Inter-SM Contention}
While elastic block design can resolve intra-SM thread-slot contention, inter-SM memory (e.g., DRAM, L2 Cache) fetching contention can still be a severe problem if blocks inside a kernel are directly launched. DNN kernels often use a large number of blocks to hide stall cycles due to data access, thus, when multiple DNN inference requests arrive in rapid succession, multiple SMs are allocated to execute the requests (e.g. memory bus) have to wait for each other, leading to decreased execution performance. 



Miriam proposes an elastic grid generator that slices the initial grid into multiple smaller grids. This approach can improve resource utilization and reduce inter-SM contention by allowing more efficient memory accesses across multiple SMs. 
Elastic grid generation implies a kernel slicing plan: Given a kernel $K$, a slicing plan $P(K)$ is a scheme that slices $K$ into a sequence of $n$ slices $[s0, s1, s2,..., s_{n-1}]$ based on thread-block-granularity partitions. 

%

Thus, given a set of kernels, the problem is to determine the optimal grid slicing policy of the initial kernel when co-running with other tasks with different workloads.
To formulate, as for a DNN kernel $K$ with $M$ thread blocks, a dichotomy algorithm-based slicing plan $S(K)$ can be applied to $K$. Specifically, there would be a sequence of slicing schemes represented as:
\begin{equation}
\begin{aligned}
S(K)=(\frac{M}{2^{n}},\frac{M}{2^{n-1}}...,M), n=\operatorname*{max}_{i} \{M\, \, mod\, \,  2^{i}=0\}
\end{aligned}
\label{limform}
\end{equation}
where $n$ is the power index of $2$ to be divided. By doing this, we enable normal kernels to be issued with a flexible number of thread blocks on SM, co-locating with critical kernels. By dividing the single kernel into multiples, the sliced grids can be scheduled to run independently by the GPU, allowing the GPU to interleave the execution of them with the execution of other critical kernels. The elastic grid design efficiently reduces co-locating kernels' inter-SM memory contention by improving the time-multiplexing potential of the kernel with other kernels, allowing the GPU to better balance the allocation of resources and maximize overall performance. 


\begin{table}[]
 \caption{GPU Architecture Parameters}
\begin{tabular}{|l|l|}
\hline
\textbf{Symbol}                                & \textbf{Parameters} \\ \hline
SM              & Streaming multiprocessors.          \\ \hline
$N_{SM}$                   & \makecell[l]{Number of streaming  multiprocessors on GPU.}          \\ \hline
$N_{blk\_{rt}}$    & \makecell[l]{Number of thread blocks in a dispatched \\critical kernel.}          \\ \hline
$N_{blk\_{be}}$                         & \makecell[l]{Number of thread blocks in a dispatched elastic \\normal kernel.}          \\ \hline
$S_{blk\_{rt}}$           & \makecell[l]{Number of launched working threads of each \\thread block in a dispatched critical kernel.}          \\ \hline
$S_{blk\_{be}}$  & \makecell[l]{Number of working threads of each thread \\block in a dispatched elastic normal kernel.}          \\ \hline
$L_{threads}$ & \makecell[l]{Limitations on the number of working threads.}          \\ \hline
\end{tabular}
\label{limitertable}
\end{table}
\subsection{Workload-balanced-guided Design Space Shrinking}\label{subsec:deisgnspace}

We need to determine the execution parameters of the elastic kernel at run-time, which includes the grid number($N_{blk\_{be}}$) and the block size($S_{blk\_{be}}$). We call each pair of execution parameters a schedule. A main challenge here is the huge number of feasible schedules, which makes it difficult to enumerate schedules or heuristically find optimal ones at run time.
The total number of feasible schedules is exponential to the number of operators in the incoming model and the size of input data. For example, an implemented \textsc{AlexNet} model in the Tango benchmark with an input image size of 3x224x224 can have up to $2.2 \times 10^{25}$ feasible schedules for all \textsc{Conv} kernels \cite{tango}. 


To address this challenge, we shrink the design space for each kernel by removing combinations of elastic grid sizes and block sizes that may result in dispatch failure due to severe resource contention. In another word, Miriam narrows down the design space by eliminating configurations that are expected to have low performance. 



When multiple kernels are co-running, thread blocks from different kernels can have many possible inter-leavings of SM-level contention or inefficiency. We propose two constraints to address these issues as shown in Eq. \ref{lim}, and the specific parameters of these factors are shown in Table 1. 

\begin{equation}
\begin{cases}
 & N_{blk\_{be}}  \leqslant  N_{SM} \ - \ N_{blk\_{rt}} ~~ mod ~~ N_{SM}\\
 & S_{blk\_{be}} \leqslant L_{threads} \ - \ blk\_{size_{rt}} 
\end{cases}
\label{lim}
\end{equation}

The first constraint is based on the observation that workload across SMs is unbalanced. This kind of imbalance appears broadly when the number of thread blocks is not a multiple of the number of SMs inside an edge GPU. To address this issue, we prune cases where the number of thread blocks of elastic kernels exceeds the remaining available SMs after dispatching all the thread blocks from critical kernels.
The second constraint addresses intra-SM workload balance, which aims to reduce contention between thread blocks from different kernels competing for resources within an SM. It is necessary to ensure that each SM has as much workload as possible and that the workload is balanced. If the workload in an SM is too light, then the resources in that SM may be wasted. On the other hand, if the workload in an SM is too heavy, it may lead to resource contention and performance degradation. We prune cases when the working threads of an elastic kernel exceed too much of the spare intra-SM resources after being occupied by blocks from the critical kernel based on the intra-SM workload balance constraint.

To formulate these two inefficiency cases, we define \textit{WIScore} as a workload imbalance metric:
\[WIScore= \frac{N_{blk\_{rt}} ~ mod ~ N_{SM}+N_{blk\_{be}}}{N_{SM}} * \frac{S_{blk\_{be}}+S_{blk\_{be}}}{L_{threads}}\eqno{(4)}\]

\noindent
where the value of \textit{WIScore} ranges from [0,1]. Another factor we consider when shrinking the design space is the dispatch overhead for the elastic kernels. To ensure that the potential schedule generated for each elastic kernel is feasible and does not violate critical decision-making requirements. Miriam prunes these cases using \textit{OScore}: 
\[
OScore = 
\begin{cases}
 &1 \ \ \sum LO_{blk}(k_{be\_{i}}) < MAX_{blk}, \forall i \in [1,N_{shard}] \\
 & and \ \ \sum LO_{pt}(k_{be\_{i}}) < MAX_{pt}, \forall i \in [1,N_{shard}] \\
 &0 \ \ Otherwise
\end{cases}
\eqno{(5)}
\]

\noindent
where function $LO()$ represents the launch overhead which equals the sum of the launching time for each elastic kernel fragment, subtracting the launching time for the initial normal kernel. \textit{OScore} is set to 0 when the overhead exceeds the maximum acceptable bar we set, which is a constant number.

The product of the $WIScore$ and $OScore$ values that are computed for each elastic kernel candidate gives a metric that can be used as a design space narrowing navigator for the performance boundary. Specifically, by multiplying these two scores ($WIScore * OScore$), we can identify the candidates that are likely to achieve the best performance within the given design space. Miriam computes it for every possible combination of elastic kernel implementation settings. Determining the optimal percentage of candidates to select is difficult since it is unclear how many candidates need to be chosen to ensure that Miriam finds the best parameters within the pruned design space. Thus, we test some representative tensor operations (such as convolution in CifarNet \cite{cifarnet} and matrix multiplication in GRU \cite{gru}) and then picks out the top 20\% combinations among all the candidates to be used in the next stage of runtime kernel coordination. Through these tests, we do not find any cases in which the model prunes the best-performing set of parameters.

With the assistance of constraint injections, we can greatly reduce the design space without sacrificing the candidate elastic kernel's performance. This feature is especially useful given the large number of possible kernel configurations in modern edge GPUs.




{\color{blue}

}

\subsection{Source-to-Source Elastic Kernel Transformer}
\begin{figure}[t!]
\centering
\includegraphics[width=0.48\textwidth]{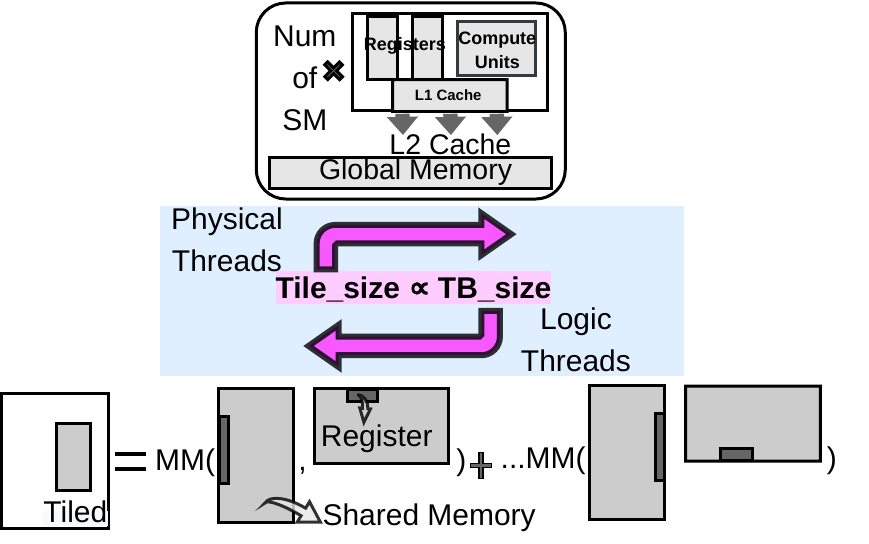}
\caption{Grid/Block size cannot be directly modified in case of recomputation/computation error.}
\label{transformer}
\end{figure}
Before assessing the effectiveness of elastic kernel design, it is crucial to investigate whether the grid or block sizes of DNN kernels can be modified directly from the original user-developed or compiler-generated GPU programs. An experiment was conducted on the benchmarks of Tango \cite{tango} to evaluate the effectiveness of direct kernel transformation. The results of the experiment showed that only 7.4\% of the implemented kernels in the Tango benchmarks were compatible with grid/block size adjustment without requiring modifications to computation schedules inside kernels. 
This is because that the block size and grid size defined in a kernel are determined by the computation schedule of the kernel: either directly written in CUDA codes or through declarative loop-oriented scheduling primitives in DNN compilers, which bind symbolic-extent logical threads with physical GPU threads, as is shown in Fig.~\ref{transformer}. This constraint motivates us to design a source-to-source kernel transformer that can support our elastic kernel design.

Miriam rapidly equivalently transforms a DNN kernel by injecting a piece of code at the beginning of each kernel, which checks the computation and memory offsets to realize where it begins and ends after being evicted. Specifically, we compute a global thread identifier and use it as a basis for SM-level workload distribution.  This identifier takes the thread ID as input and produces a corresponding index for the data element accessed by the thread. We replace references regarding physical threads (e.g. $GridDim$) and identity variables (e.g. $threadIdx.x$) in the original kernel codes with logical equivalents. Miriam employs two approaches for implementing the index function: computation-based and memory-based. The computation-based approach computes the index within the kernel when the thread accesses the corresponding data element. Alternatively, in the memory-based approach, the indices are pre-calculated on the host side (i.e., the CPU) prior to kernel launch and stored in shared memory for use during kernel execution.

\section{Runtime Dynamic Kernel Coordination}
This section introduces our design for the online scheduler of elastic kernel coordination. First, we call each elastic kernel (i.e., elastic grid and elastic block) as \textit{elastic kernel shard}. Our guidelines for designing the coordinator are two-fold: maximizing overall real-time performance and mitigating resource contention. To achieve these goals, our runtime coordinator constantly monitors the available GPU resources, both from the critical kernels and elastic kernels. It then determines which elastic kernel shards can co-run effectively with the critical kernels.

\noindent\textbf{Execution timeline of co-running kernels.} Upon receiving multiple normal task requests $b1...bn$, Miriam pushes all the kernels into a normal tasks queue and the kernels are dispatched to the GPU semantic through multiple streams. Once a critical task arrives, Miriam will instantly select appropriate elastic kernel fragments of the following normal kernel in a "bin-packing" manner, considering the current intra- and inter-SM-level resource distributions. After that, once the critical kernels finished executing, all the kernels from normal tasks will re-occupying the GPU. 



\noindent\textbf{Grid/block size determination of elastic kernels. }
During runtime, a fixed size for elastic grids and block settings for elastic kernels can easily become inefficient with the optimal co-scheduled elastic kernel shards varying with different co-running with critical kernels. For example, if one critical kernel finishes and there still exists half of the computations unfinished from the co-locating elastic kernel, the rest half of thread blocks from it lead to severe resource contention or under-utilization when co-locating with the subsequent critical kernel. The selection policy for elastic kernel shards is crucial in order to prevent latency interference with critical tasks. To ensure optimal performance, one approach is to build a duration prediction model for the formation of operator groups based on runtime performance events (e.g. cache misses and global memory bandwidth)\cite{orion, horus}, and control the kernel overlap based on the model. However, runtime events are not supported on edge GPUs like Nvidia Jetson devices, and the hardware events reported by tools like Nsight Sys and Nsight Compute can only be obtained with high overhead. Thus, this method cannot be applied to our problem (kernel overlaps are not determined) in a practical way.


\begin{figure}[t!]
\centering
\includegraphics[scale=0.9]{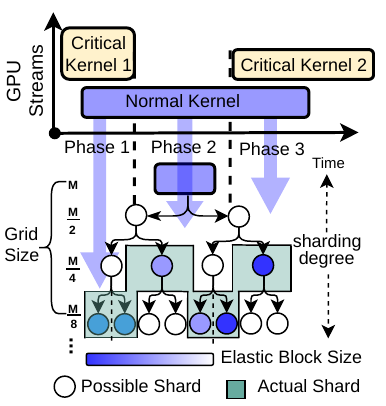}
\caption{Shaded Binary Tree Construction for Kernel Shards Formations. ES refers to the elastic kernel, and EBS refers to the elastic block size. The sharding degree represents the degree of elastic kernel splitting depth.}
\label{binary}
\end{figure}
To address these challenges, Miriam adopts a greedy scheduling policy. Specifically, when the elastic kernel partially overlaps with the critical kernel, the kernel coordinator must carefully balance the resources allocated to each kernel. In this case, the coordinator needs to ensure that the padded elastic kernel does not interfere with the execution of the critical kernel, while still using as many available resources as possible. When the padded kernel runs on its own, the kernel coordinator can allocate all of the available resources to the kernel, since there are no other tasks running on the GPU. This allows the kernel to run as efficiently as possible, without any interference from other tasks. To efficiently manage these elastic kernels while achieving the goal, we propose a dynamic-sized shade binary tree approach for elastic kernel shards formation to achieve high runtime efficiency and low resource contention from different combinations of overlapped kernels.

Our shaded binary tree structure is an abstract for managing the elastic kernel shards, which is similar to a complete binary tree structure of shards, as is shown in Fig. \ref{binary}. The root of the tree represents the kernel from the normal tasks, whose initial grid size is $M$. Each node corresponds to a part of computations, or potential thread blocks to be dispatched inside the kernel. The shading property for each node is the elastic block size of the thread block. Directed edges indicate the potential sliced peers for the unfinished computations left over from the predecessor. The whole structure is composed of the actual shard and the virtual shard. The actual shards are the ultimately formed elastic kernel shards that are to be dispatched, and the virtual shards are the potential fragments of the elastic kernel that would not be dispatched.

Miriam relies on the dynamic shaded kernel binary tree structure to manipulate the elastic kernels from normal tasks and determines the elastic kernel shards with heuristics based on the number of thread blocks of kernels from both critical and normal tasks. As is shown in Fig. \ref{binary}, which illustrates the life cycle of an elastic normal kernel. For elastic fragment selection from normal kernels, the policy is to pick a set of elastic blocks from the head of the shaded kernel binary tree to share SM-level resources with co-locating thread blocks from resident critical kernels with trivial contention. Miriam proposes to utilize a policy to ensure that the elastic blocks from normal kernels will only use the left-over resources from the critical kernels.


  


    

\section{Evaluations} \label{evaluation}

\subsection{Experiment Setup}

We implemented Miriam based on NVIDIA CUDA 11.2 \cite{cuda} for elastic kernel generation and online kernel scheduling, and Python3.6 for the source-to-source kernel transformer. 
\vspace{-0.6em}
\subsubsection{\textbf{Implementation and Testbed.}}
Our experiments are conducted on an NVIDIA GeForce RTX 2060 that features 1920 CUDA cores and an NVIDIA Jetson AGX Xaiver with Pascal GPU architecture with 256 NVIDIA
CUDA cores \cite{cuda}. We implemented Miriam with NVIDIA CUDA 11.2 for elastic kernel generations and Python3.6 for the end-to-end kernel transformation. Note that Miriam is extensible and can work well on other GPU platforms that officially support OpenCL, HIP or other CUDA alike programming paradigms such as AMD Embedded Radeon™ E9170 \cite{amd}.
\vspace{-0.6em}
\subsubsection{\textbf{DNN Workloads.}}
We use six popular DNN models from both computer vision and language processing fields to evaluate Miriam. Inspired by DISB \cite{reef}, we build a benchmark named MDTB (Mixed-critical DNN Task Benchmarks) based on both CUDA implemented Kernels to fully demonstrate the performance and generalization of our framework, summarized in Table \ref{mdtb}. MDTB benchmark simulates three patterns for inference tasks from user requests: (1). Arrival in uniform distribution. The client sends inference requests at a fixed frequency (e.g. 10 requests/second), which simulates critical applications such as pose estimation. (2). Arrival in Poisson distribution, which simulates event-driven applications such as obstacle detection. (3). Closed-loop workloads simulate when the client keeps sending inference requests. 

We choose five representative DNN models in MDTB, including AlexNet \cite{alexnet}, SqueezeNet \cite{squeezenet}, GRU \cite{gru}, LSTM \cite{LSTM}, ResNet \cite{resnet}, and CifarNet \cite{cifarnet}, all implemented in CUDA. We conduct neural network inference with a 224x224x3 single batch of images as the input to mimic the inference in real applications.

\begin{table}[]\Huge
\resizebox{0.46\textwidth}{!}{
\begin{tabular}{c|cccc}
\hline
\textbf{MDTB}                                                                       & \textbf{A}                                                                     & \textbf{B}                                                                        & \textbf{C}                                                            & \textbf{D}                                                                  \\ \hline
\begin{tabular}[c]{@{}c@{}}Critical Task \\ Frequency (req/s)\end{tabular} & \begin{tabular}[c]{@{}c@{}}AlexNet\\ Closed-loop \end{tabular} & \begin{tabular}[c]{@{}c@{}}SqueezeNet\\  Uniform (10 reqs/s)\end{tabular} & \begin{tabular}[c]{@{}c@{}}GRU\\ Poisson (10 reqs/s)\end{tabular}    & \begin{tabular}[c]{@{}c@{}}LSTM\\ Uniform (10 reqs/s)\end{tabular} \\ \hline
\begin{tabular}[c]{@{}c@{}}Normal Tasks\\ Frequency (req/s)\end{tabular}   & \begin{tabular}[c]{@{}c@{}}CifarNet\\ Closed-loop\end{tabular}        & \begin{tabular}[c]{@{}c@{}}AlexNet\\ Closed-loop\end{tabular}            & \begin{tabular}[c]{@{}c@{}}ResNet\\ Closed-loop\end{tabular} & \begin{tabular}[c]{@{}c@{}}SqueezeNet\\ Closed-loop\end{tabular}   \\ \hline
\end{tabular}
}
\caption{MDTB Workload Description.{\color{blue}}}
\label{mdtb}
\end{table}
\vspace{-0.6em}
\subsubsection{\textbf{Baselines.}}
We compare Miriam with multiple DNN scheduling approaches on edge GPU. \textbf{Sequential} selects one model from both task queues (critical and normal) in a round-robin fashion and performs the inference one by one. In this mode, the critical tasks run independently, occupy the GPU resources, and can have optimal end-to-end latency for critical tasks. \textbf{GPU Multi-stream with Priority}  enqueues kernels from both critical and normal tasks at the same time, and models are executed in parallel. This is adopted by NVIDIA Triton \cite{Triton}. \textbf{Inter-stream Barrier (IB)} is the state-of-art multi-DNN operator scheduling method based on multi-stream \cite{automatediccad}. It uses inter-stream barriers to manually synchronize kernel dispatch among different kernels. In this mode, the concurrency among kernels can be controlled by utilizing stream and synchronization-based mechanisms.
\vspace{-0.6em}
\subsubsection{\textbf{Metrics.}}
We use the overall throughput, the end-to-end latency for critical tasks, and the achieved occupancy as our evaluation metrics.
  
\noindent\textbf{End-to-end Latency of Critical Tasks.} This metric measures the end-to-end inference speed of critical tasks with real-time demands. 

\noindent\textbf{Overall Throughput.} This metric represents how many requests from users can Miriam serve on the target edge GPU. 

\noindent\textbf{Achieved Occupancy.} By definition, achieved occupancy is the average ratio of active warps on an SM to the maximum number of active warps supported by the SM\cite{cuda}, defined as below:
\vspace{-0.2em}
\begin{equation}
\begin{aligned}
Achieved\  Occupancy = \frac{Active\_warps / Active\_cyles}{MAX\_warps\_per\_SM} \nonumber
\end{aligned}
\label{7}
\end{equation}
\vspace{-0.2em}
We use this metric to evaluate the fine-grained GPU utilization of our system performance.

\begin{figure*}[!ht]
\centering
\includegraphics[width=1\textwidth]{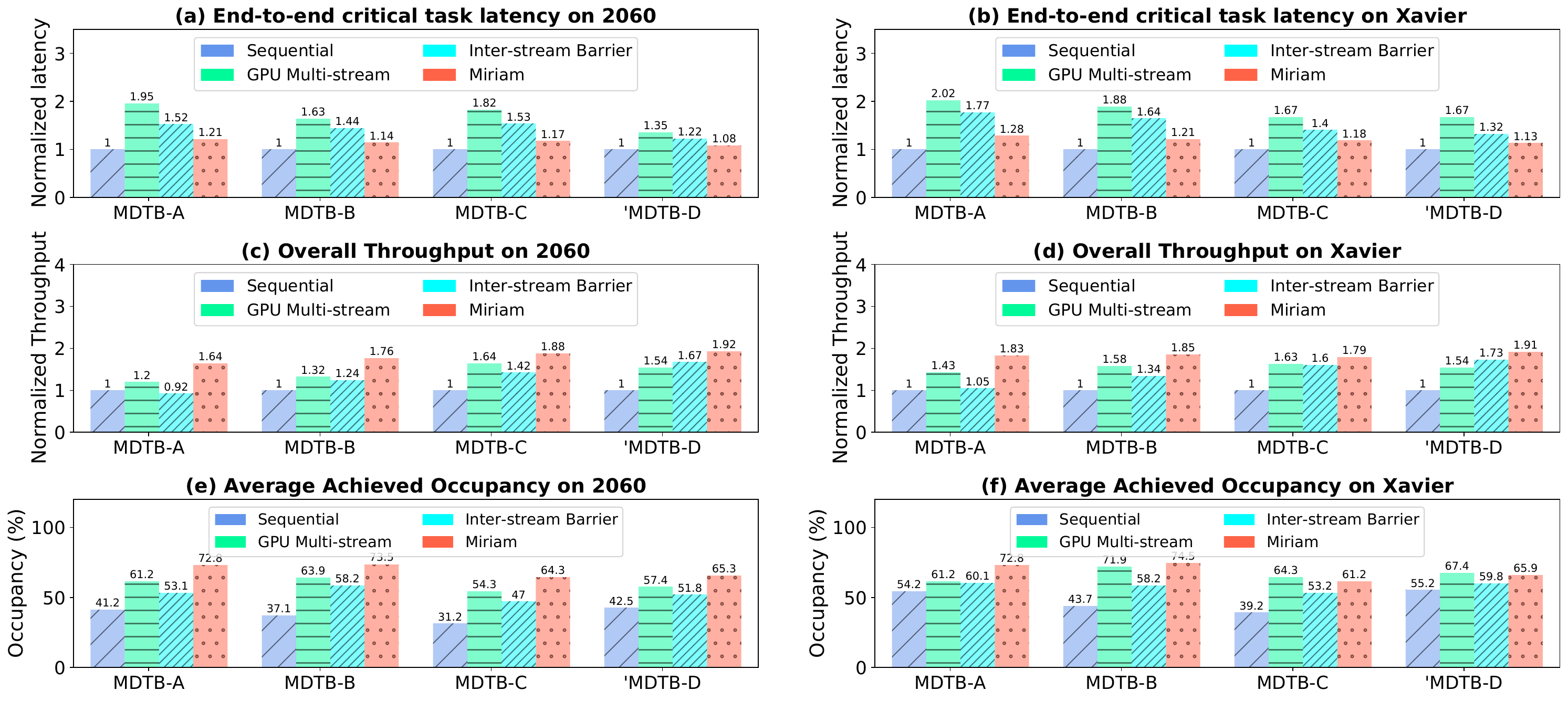}
\caption{Comparison of end-to-end real-time task latency, overall throughput (including both critical and normal tasks), and average achieved occupancy among different GPU scheduling approaches.}
\label{mainresult}
\end{figure*}

\subsection{Overall Performance}
To reflect the performance gain of system overall throughput with little sacrifice on the real-time performance of the critical tasks, we compare Miriam against other GPU scheduling approaches under MDTB A-D workloads on two edge GPU platforms.
We merge discussion of the uniform distribution and poisson distribution of critical task requests because their workloads are comparable. This allows us to analyze and discuss their similarities more efficiently.

\noindent\textbf{Closed-loop Critical Tasks (MDTB A).} Workloads with closed-loop critical tasks (AlexNet) experience significant resource contention when co-running with normal tasks (CifarNet). Fig. \ref{mainresult} (a)-(d) show that: compared to Sequential, Multi-stream and IB increase the critical task latency by 1.95$\times$ and 1.52$\times$ on 2060 and 2.02$\times$ and 1.77$\times$ on Xavier, respectively, while Miriam incurs only a 21\% and 28\% overhead on critical tasks. Miriam also improves overall throughput by 64\% and 83\% on the two platforms, outperforming other approaches significantly under MDTB A workloads. We observed that IB's throughput performance is even worse than Sequential's due to the frequent launching of critical tasks require the insertion of more synchronization barriers among GPU streams to manage kernel groups, resulting in significant overhead. In terms of achieved occupancy, Fig. \ref{mainresult} (e) and (f) demonstrate that Miriam leads to higher SM-level GPU resources compared to other baselines. It is important to note that achieving nearly 100\% theoretical occupancy is difficult for DNN inference tasks due to their large thread blocks, which can easily lead to resource idleness or SM incapacity to cover memory access latency \cite{deepcuts}.

\noindent\textbf{Uniform/Poisson Critical Tasks (MDTB B, C, and D).} As the launching frequency of critical workloads decreases, the overall throughput of all approaches improves with different degrees compared to vanilla Sequential due to increased opportunities for normal tasks to share GPU resources with critical tasks. We observed that Miriam outperforms other approaches in this scenario. For instance, using MDTB B, C, and D on Xavier, Miriam increases overall throughput by 1.85$\times$, 1.79$\times$, and 1.91$\times$ over Sequential, which is much better than the other baselines. While both Multi-stream and IB also yield improved throughput compared to Sequential with 1.34$\times$~1.73$\times$, they lead to severe latency degradation for the critical tasks by 32\%~88\%, whereas Miriam only incurs a latency overhead of less than 21\% for these benchmarks. This improvement can be attributed to our elastic kernel design and runtime dynamic kernel coordination approach. Since the Sequential approach exhibits the shortest latency for each critical task, our comparison demonstrates that Miriam maximizes overall throughput while preserving the end-to-end latency of critical tasks. From a GPU utilization standpoint, Miriam increases the average active warps of each cycle, resulting in better SM utilization. These results confirm the effectiveness of our elastic kernel sharding approach and demonstrate our ability to effectively pad critical kernels.

We observe that the performance improvements offered by Miriam may not always result in higher SM occupancy on Jetson Xavier. This is because Xavier has much fewer onboard resources and a smaller number of SM compared to 2060. Additionally, the relatively low memory bandwidth of the Xavier can limit the amount of data that can be transferred between the memory and SMs, leading to performance bottlenecks with complex models. The thermal design power of the Xavier is also relatively low compared to 2060, which can limit the amount of power that can be consumed by the GPU and the amount of heat that can be generated. This can negatively impact the clock speed of the processor cores and the amount of parallelism that can be achieved, which in turn can have a negative impact on the relationship between SM occupancy and performance. \\

\begin{figure}[t!]
\centering
\includegraphics[width=0.45\textwidth]{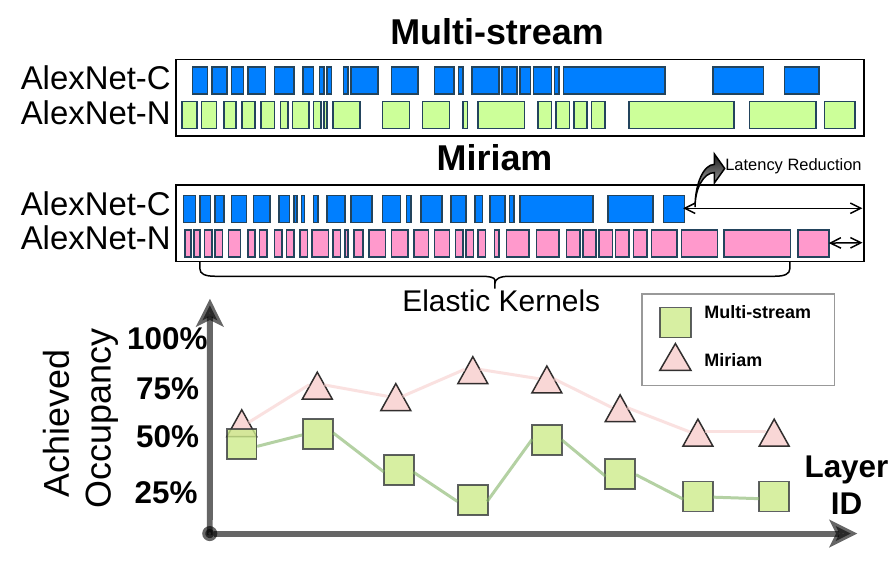}
\caption{(Upper) The active kernel-level timeliness of two co-running AlexNet models with mixed-criticality, which is profiled from the Nsight System. (Lower) The average achieved occupancy for each layer of the critical AlexNet.}
\label{alex}
\end{figure}

\vspace{-1em}
\subsection{In-depth Analysis of Miriam }
To better understand why Miriam performs better than other GPU scheduling approaches under severe contention circumstances, we provide a in-depth analysis in this section, with two AlexNet models co-running on a single 2060 GPU named AlexNet-C which serves as the critical task, and AlexNet-N which serves as the normal task. Both tasks are launched in a closed-loop manner. 

In Fig. ~\ref{alex}, the upper two rows show the timelines of active kernels from the two co-running DNN tasks, which demonstrate the performance difference between Miriam and Multi-stream. The figure is sketched based on real profiling results achieved from NVIDIA Nsight Sys \cite{nsys}, in which we use the blue color to represent the critical task, green color to represent normal tasks launched by vanilla Multi-stream, and pink color represents elastic kernels of the normal task by Miriam. As shown in the figure, there are obviously more pink blocks than green blocks, and these pink blocks are tightly padded with the blue blocks, which can be a showcase of the elastic kernel shards padded with the critical kernels. The end-to-end latency of AlexNet-C in Miriam is much lower than that in Multi-stream. 

We also show the corresponding achieved occupancy of this case in Fig. ~\ref{alex}. The average layer-wise achieved occupancy for Miriam is 65.25\% and is 32.9\% for Multi-stream. As mentioned, more average active warps per cycle and less contention overhead is the key to improving the parallelism while preserving the speed of critical tasks.

\subsection{Evaluations on Design Space Shrinking}
\begin{figure}[t!]
\centering
\includegraphics[width=0.43\textwidth]{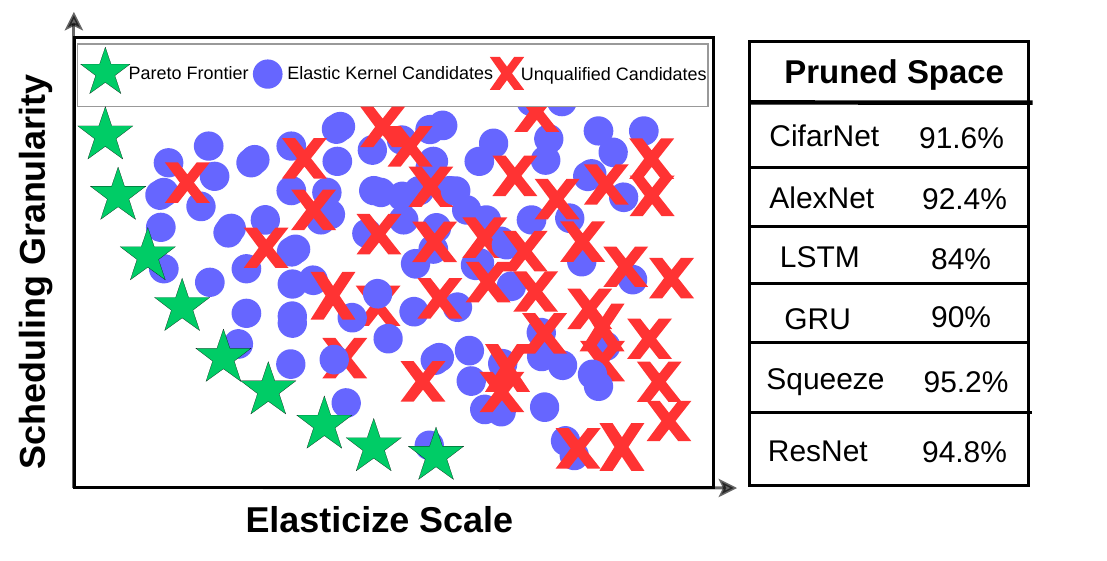}
\caption{Shrinking the design space of elastic candidates for different DNN Models. Miriam picks up elastic kernels lying in the Pareto Frontier (for visualization) of trade-off space between the elasticized scale and the scheduling granularity.}
\label{shrink}
\end{figure}
\begin{figure*}[t!]
\centering
\includegraphics[width=1\textwidth]{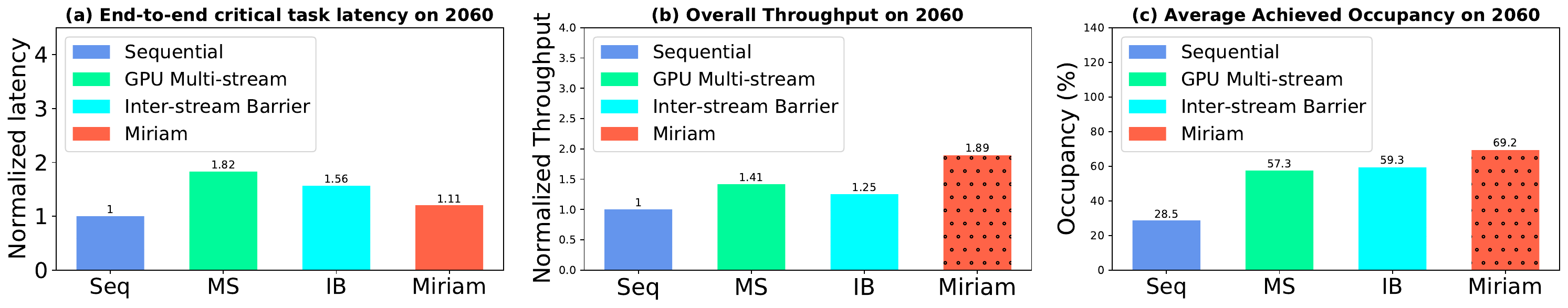}
\caption{Comparison of end-to-end real-time task latency, overall throughput, and average achieved occupancy using different scheduling schemes with our LGSVL simulated workloads.}
\label{realtraceresult}
\end{figure*}

\begin{figure}[t!]
\centering
\includegraphics[width=0.45\textwidth]{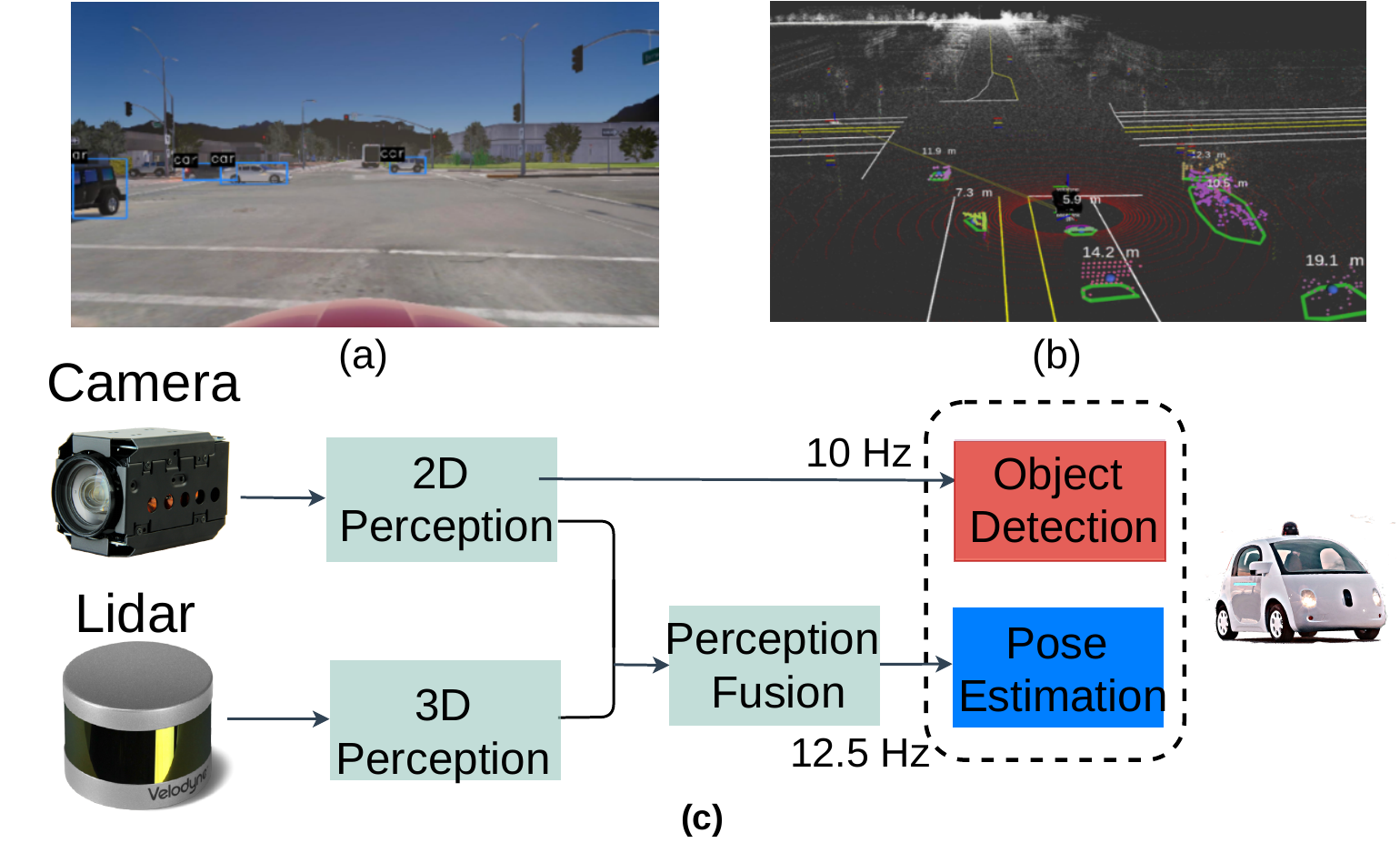}
\caption{Real-world trace collected from LGSVL simulator, where (a) is the object detection result based on image data, (b) is the result with lidar point cloud data, and (c) depicts the setting of our collected trace.}
\label{realtrace}
\end{figure}

Miriam filters out the definitely-slow cases (80\%) by applying hardware limiters, as detailed in Chapter 6.3. The trade-off between elasticized scale (i.e., the dynamic shaded binary tree's depth, as discussed in Chapter 7) and scheduling granularity is a critical consideration for different implementations of elastic kernels, as shown in Fig. \ref{shrink} to guide the further shrinking process. For instance, an elastic kernel shard with $elastic_grid_size=1$ is flexible to accommodate other critical kernels, but launching overhead for such a shard may be too large due to the increased number of kernel shards. Fig. \ref{shrink} summarizes the pruned space of candidate elastic kernels from the models in MDTB, ranging from 84\% to 95.2\%. The expected pruned space may differ across candidate models due to multiple factors, such as the complexity of the models (i.e., the operator types used) and the input size.

\subsection{Case Study: Autonomous Driving with LGSVL}
We further use a real-world trace from an open autonomous driving platform (i.e., LG SVL \cite{LGSVL}) as the workload, which provides a realistic arrival distribution of critical tasks (i.e., obstacle detection) and normal tasks (i.e., pose estimation) in autonomous driving. 

The trace was collected from a 3D Lidar perception module and a 2D camera perception module when running the LGSVL simulator, and we selected backbones from the models included in our MDTB benchmark, they are SqueezeNet for simulation of pose estimation as the normal task (lidar data), and ResNet for obstacle detection as the critical task (camera data). The clients send the inference requests in a uniform distribution, with 12.5 Hz frequency for the normal task and 10 Hz for the critical task, as is shown in Fig. \ref{realtrace}. The experiment was conducted on GTX 2060. 

Fig.~\ref{realtraceresult} demonstrates the experimental results for this real-world workload. Compared to Sequential, Multi-stream and IB increase the overall throughput by 1.41$\times$ and 1.25$\times$, while amplifying the critical task latency by 82\% and 56\%, respectively. Due to the low launching frequency of both critical and normal tasks (10 and 12.5 Hz), the elastic kernels of the normal task can execute concurrently with the critical task with little eviction overhead for elastic kernel shards. Finally, Miriam achieves 89\% improvement of overall throughput compared to Sequential, and only incurs 11\% latency overhead for the critical task. This proves how Miriam can achieve large improvement of throughput based on our elastic kernel design with little sacrifice on critical task latency, which is also confirmed by our high SM occupancy among all baselines shown in Fig. \ref{realtrace} (c).
\vspace{-1em}
\subsection{System Overhead}
The scheduling overhead of Miriam mainly consists of two parts. The first part is the runtime elastic kernel shards selection, which scans the shard candidates and has the complexity of $O(N)$. Owing to the low complexity of the scheduling mechanism in Miriam, we find that their overall average overhead for serving in each DNN model is less than 0.35 ms. The second part is the launch time overhead for critical kernels due to the padding of the elastic kernels, we evaluate this overhead and found that in most (over 80\%) cases, the overhead is less than 15 us. This latency overhead is mainly because of contention on the texture cache and L2 memory, which we leave for future work. 

\vspace{-1em}
\section{Discussion} \label{discussion}

\noindent\textbf{Scalability.} We believe that Miriam has the potential to be scaled beyond pair-wise DNN tasks co-running and can support more general tasks. However, due to the large number of co-running kernel possibilities, some additional considerations must be taken into account. These include establishing a scheduling policy for normal tasks with the same priority, as well as finding an efficient way to perform offline kernel profiling since the design space increases exponentially. 

\noindent\textbf{Integrated with DNN Compiler.} Representative DNN compilers like TVM \cite{TVM} can generate high-performance DNN kernels with low latency using auto-tuning \cite{ansor}. However, DNN compiling is an offline approach with a long compilation time, and the generated kernels can not be easily modified at runtime. This creates a gap between static compilation and dynamic scenarios in IoT applications, particularly when on-device resources become available dynamically.
To fill this gap, Miriam can serve as a post-compiling runtime to ensure that the on-device resources are fully utilized during runtime in an adaptive manner. 

\noindent\textbf{Orthogonal to Other Approaches.} Miriam can work symbiotically with other optimized DNN execution approaches, such as model compression \cite{rt-mdl}, and edge-cloud offloading \cite{edgeml}, to execute multi-DNN workloads effectively.With such a collaborative approach, it becomes possible to achieve improved runtime performance and better resource utilization, enabling effective execution of multi-DNN workloads in resource-constrained edge computing environments.






\vspace{-1.2em}
\section{Conclusion}
We propose a novel system named Miriam that addresses latency and throughput problems of co-running multiple DNN inference tasks on edge GPUs. The proposed system utilizes elastic kernels to facilitate fine-grained GPU resource re-mapping and a runtime dynamic kernel coordinator to support dynamic multi-DNN inference tasks. Experimental results on a benchmark we built on two types of edge GPU show that Miriam can significantly improve the overall system throughput while incurring minimal latency overhead for critical tasks, compared to dedicating the GPU to critical tasks.
\clearpage
\bibliographystyle{plain}
\bibliography{references}


\end{document}